\def\year{2018}\relax
\documentclass[letterpaper]{article} 
\usepackage{aaai18}  
\usepackage{times}  
\usepackage{helvet}  
\usepackage{courier}  
\usepackage{url}  
\usepackage{graphicx}  
\usepackage{booktabs} 
\usepackage{multirow}
\usepackage{array}
\usepackage{amsmath}
\usepackage{amsfonts}
\usepackage{xcolor}

\usepackage{soul}

\begin{document}
%
\title{From the User to the Medium: Neural Profiling Across Web Communities}
\author{Mohammad Akbari, Kunal Relia, Anas Elghafari, Rumi Chunara\\
New York University\\
\{akbari,krelia,anas.elghafari,rumi.chunara\}@nyu.edu\\
}
\maketitle
\begin{abstract}
Online communities provide a unique way for individuals to access information from those in similar circumstances, which can be critical for health conditions that require daily and personalized management. As these groups and topics often arise organically, identifying the types of topics discussed is necessary to understand their needs. As well, these communities and people in them can be quite diverse, and existing community detection methods have not been extended towards evaluating these heterogeneities. 
This has been limited as community detection methodologies have not focused on community detection based on semantic relations between textual features of the user-generated content. Thus here we develop an approach, NeuroCom, that optimally finds dense groups of users as communities in a latent space inferred by neural representation of published contents of users. By embedding of words and messages, we show that NeuroCom demonstrates improved clustering and identifies more nuanced discussion topics in contrast to other common unsupervised learning approaches.

\end{abstract}

Online communities are places where individuals have found support and places to exchange customized disease-specific information~\cite{frost2008social}. 
For non-communicable diseases such as diabetes, social platforms have become very relevant places where individuals connect to learn about their condition outside of clinical settings. 
This is important as diabetes manifests in an evolving and heterogeneous manner, shifting in concert with population-wide alterations in behavioral and lifestyle factors and disease management strategies \cite{weitzman2011sharing}. 
Decades of studies have shown that risk of diabetes can differ by ethnic or gender groups and as well efficacy of interventions can vary by population subgroups \cite{sarkar2006self}. Thus individuals find the personal experiences of others managing their same conditions useful for learning about new efficacious interventions, or other day-to-day strategies. 
As the data in social media and web groups is generated by individuals in  unstructured formats and venues, and is constantly updated and changing, there is a need for methods to distill and extract the types of topics and groups being discussed. 

Thus in this paper, we harness the increasing use of neural representations and statistical natural language processing to demonstrate an approach for embedding content of users' posts and discovering communities in the same space based on the \emph{content} of online posts. Specifically, we adopt neural text representation to model the semantic link amongst words in a lower dimensional space and then perform community detection and profiling in that space which maintains a level of topicality in discovered user communities, messages and profiles. 
We make the following contributions:

\begin{itemize}
\item We demonstrate a neural approach (NeuroCom) for learning a low-dimensional latent space from the embedding of users' posts, which enables community detection. 
\item We combine the neural framework with inference and qualitative methods to demonstrate how it can be used to learn and compare the substantive topics within and between diabetes communities from several large scale real-world datasets. 
\end{itemize}

\section{Related Work}

Methods for identifying relevant groups is an active area of research;
a great deal of work is on graph-based data such as social or information networks. Community detection is then based on choosing an objective function that captures the  intuition of a community as a set of nodes with better internal connectivity than external connectivity \cite{leskovec2010empirical}. 
This is a rich area of research and briefly summarizing, there is work across spectral algorithms \cite{kannan2004clusterings}, measures of centrality~\cite{newman2004finding} and matrix factorization \cite{akbari2017leveraging}. 

Neural representation has been implemented for social media/short messages albeit in different ways than we propose \cite{abdelbary2014utilizing}. Gaussian Restricted Boltzmann Machines (RBM) have been used for modeling user's posts within a social network to identify their topics of interest, and finally construct communities \cite{abdelbary2014utilizing}. However,  a parametric approach was used in which it was necessary to specify the number of clusters/communities. 
In addition to not requiring such initialization, this approach is conceptually different than our proposed work in that it directly maps individuals to communities (instead of mapping the content of their posts, which may better capture heterogeneous community memberships). Further, content-based density approachefs as proposed, versus parametric ones could potentially learn a more organic number of communities. Given this gap, and the fact that content-based community detection (opposed to graph-based) may be more pertinent in health-related communities, here we explore content-based clustering of health communities.

\section{Methodology}
\begin{figure}[b]
\centering
\includegraphics[width=1.0\columnwidth]{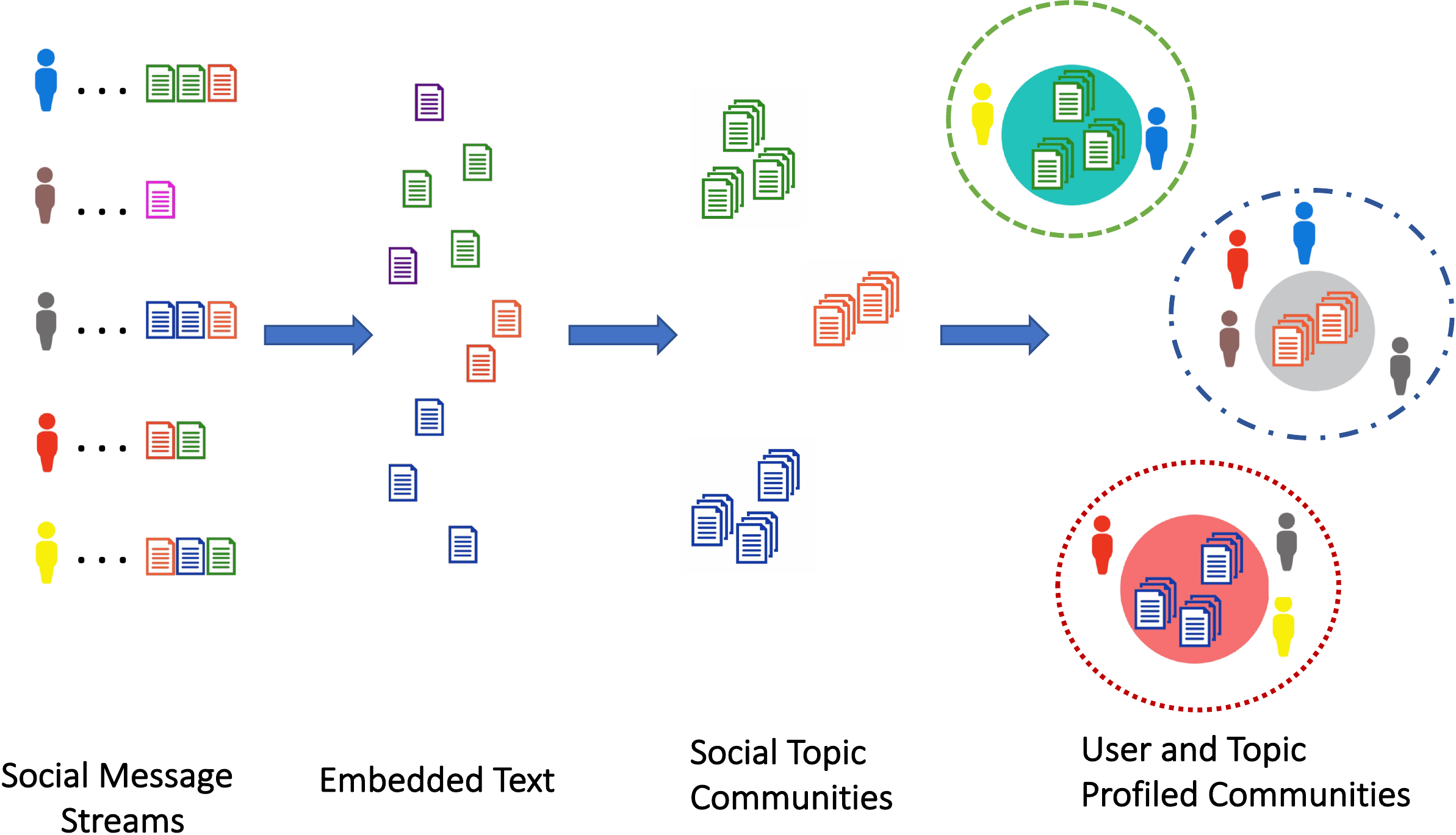}
\caption{NeuroCom framework for embedding messages, users, and communities in the same latent space.}
\label{fig::framework}
\end{figure}

In this paper, we aim at learning the representation of user posts from web and social platforms using an effective neural model. Our framework, in contrast to conventional models which just embed \emph{users} in a latent space, learns the representation of users, messages and communities in the same latent space, as shown in Figure~\ref{fig::framework}. This helps in the identification of communities that may be under-represented and thus missed in a single user's messages.

\subsection{Distributed Representation of Social Messages}

Our model is derived from a neural model, continuous BoW (\textbf{C-BoW}) \cite{mikolov2013distributed}, with a few innovations to learn the embedding of social messages.
The \textbf{C-BoW} model is a simplified neural model without any non-linear hidden layers, which learns a distributed representation for each word $w_t$ while taking care of the semantic similarity of words. 
More specifically, given a large training corpus represented as a sequence of words $w_1, ..., w_N$ , the objective of embedding is to maximize the log-likelihood, i.e. 
$\sum_{t=1}^{N}\sum_{c \in C_t} \log p(w_t \vert w_c),$
where $C_t$, referred as context word, is the set of words surrounding $w_t$, i.e., target word.

 We extend this model to learn the embedding of \emph{social messages} from the embedding of their compositional components. Formally, we based our model on the assumption that the message embedding is the average of embeddings of its n-gram constitutional elements. 
 Formally, the embedding for a social message $m$ can be stated as:
 \small
\begin{align}
\label{eq:message-similarities}
\mathbf{v}_m = \frac{1}{\vert E(m)\vert} \mathbf{V} \mathcal{I}_{E(m)}=\frac{1}{\vert E(m)\vert} \sum_{w \in \mathcal{E}(m)} \mathbf{v}_w, 
\end{align}
\normalsize
where $\mathbf{V} \in \mathbb{R}^{N \times v}$ is a matrix collecting the learned embeddings for words, $E(m)$ denotes the list of n-grams in the message $m$ and $\mathcal{I}_m \in \{ 0, 1\}^{N}$ is an indicator vector representing the compositional elements of $m$.
We adopted negative sampling framework for learning this model.

\subsection{Discovering Topical Communities of Messages}

To discover topical focuses of users, we leverage density-based clustering of social messages in latent space as it does not require the input of a predefined number of clusters, and can form clusters with arbitrary shapes~\cite{ertoz2003finding}. More specifically, we used DBSCAN which is a popular density-based clustering algorithm. 

\subsection{Topical Profiling}

To profile topics, we can compute the embedding of each community by averaging the embeddings of its involving messages and the affiliation of each user to the community by his contribution in the community messages. More specifically, the discovered clusters in the prior section are considered topical communities of users and the affiliation matrix of users is defined as, $
	\mathbf{H}_{i,j} = \frac{\mathbf{U}_{i,j}}{\sum_{j=1}^{K} \mathbf{U}_{i,j}},
$
where $\mathbf{H}_{i,j}$ denotes membership affiliation of the $i$-th user to the $j$-th community and $\mathbf{U}_{i,j}$ the message number belonging to $i$-th user in the $j$-th community.

\subsection{Demographic Profiling}
In order to better understand these communities and the users who participate in them, we infer basic demographics of participants in both the social media and on-line forum groups. The epidemiological literature indicates that any endeavor is incomplete without an understanding of the target population \cite{chunara2017denominator}, and this becomes more pertinent when working with observational and Internet-based datasets such as the work here. 

In order to accomplish this, we use inference methods which have been used in other social media-based studies \cite{huang2018cscw}. 
For age, we follow the approach in \cite{sloan2015tweets,huang2018cscw}. We classify users as under 30, or 30 and older, which, in diabetes research, has been used to qualify ``young age at onset''. 
For gender, look-up tables have been used based on user handles \cite{mullen2015gender}. However, as we did not want to introduce uncertainty and more than half of the names were not linked to a gender, we did not assess any gender composition.

\section{Experiments}

\subsection{Datasets} 
For evaluation, we chose three datasets that we anticipate will be slightly different in their content and included users.
The statistics of our datasets are shown in Table~\ref{tbl::dataset-statistics}.

\begin{table}
	\centering
    \small
	\caption{The statistics of our datasets.}
	\label{tbl::dataset-statistics}
    \vspace{0.2em}
	\begin{tabular}{|c|c|c|c|}
		\hline 
		& \# of Users & \# of Posts & \# of Words \\ 
		\hline \hline
			Support Group	& $2,927$ &  $2,773,320$ & $52,309,657$ \\ 
		\hline 
		BGnow 	& $2,888$  & $82,306$ & $1,518,384$  \\ 
		\hline 
		TuDiabetes	& $16,331$  & $554,333$  & $70,199,144$ \\ 
		\hline 
	\end{tabular} 
\end{table}

\emph{\textbf{Diabetes Support Group}}:
This dataset is collected from posts of users who follow and participate in diabetes support groups like ``diabeteslife'' or ``diabetesconnect'' on Twitter. To construct the dataset, we first gathered a set of users who followed these diabetes support groups in Twitter. 
We then crawled the Twitter timelines of these users. We selected $5$ different support groups to avoid the bias coming from a specific support group~\footnote{The following support groups were selected as seed accounts in Twitter: ``diabeteslife'', ``diabetesconnect'', ``American Diabetes Association'', ``DiabetesHealth'', and ``Diabetes Hand Foundations''}. 

\emph{\textbf{BGnow Dataset}}: 
Another dataset derived from diabetic users who actively share their wellness data on Twitter. These users not only post about their lifestyle and activities such as their diet, but also share their health information in terms of medical events and measurements like their blood glucose value~\cite{akbari2016tweets}. 
Users in this dataset are majority diabetic type $I$ and they used ``\#Bgnow'' hashtag to report their blood glucose value on Twitter.

\emph{\textbf{TuDiabetes Forum}}:
We aslo collected a dataset from the TuDiabetes forum, a popular diabetes community operated by the Diabetes Hands Foundation. It provides a rich community experience for people with interest in diabetes, including a social network, personal pages and blogs. The TuDiabetes forums are very popular, and have been active for years thus encompassing many topics.

\subsection{Baselines and Metrics}
We compare the proposed method with following baselines: \textbf{\textit{KMeans}} (a widely used clustering method in social networks~\cite{kmeancommunity} with Tf-IDF representation for users and cosine measure for similarity computation), \textbf{\textit{KMeans-Lat}} (similar to the \textbf{\textit{KMeans}} approach, however clustering is performed in a latent space derived by Eq.(\ref{eq:message-similarities})), \textbf{\textit{Biterm-LDA}} (Latent Dirichlet Allocation (LDA) topic modeling tuned for short messages), and \textbf{\textit{RBM}} (a deep model for community detection in~\cite{abdelbary2014utilizing}). We utilized Silhouette (\textbf{sil}) and normalized mutual information (\textbf{nmi}) metrics for benchmarking.

\subsection{On Quantitative Comparisons of the Model}

Table~\ref{tbl::quantitative-model-comparison} shows the clustering results of different methods in terms of \textbf{sil} and \textbf{nmi}. We followed previous research studies to tune the parameters for all baseline methods. For \textbf{Biterm-LDA}, as proposed by~\cite{yan2013biterm}, the parameters $\alpha$ and $\beta$ have been fixed to $50/K$ and $0.01$, respectively, where $K$ is the number of clusters/topics computed using grid-search. In the \textbf{RBM} model, we tuned the number of hidden units to $250$ and we then examined different number of community detection units ${25, 50, 75, 100}$ and reported the best results, as suggested in ~\cite{abdelbary2014utilizing}.

From the table, the following points can be observed: (1) \textbf{Kmeans} and \textbf{Kmeans-Lat} achieve the lowest performance in terms of both quality and consensus metrics. This is mainly attributed to the fact that \textbf{KMeans} with Tf-Idf features fails to capture similarities between words/terms in posts. (2) \textbf{RBM} and \textbf{BiTerm-LDA} outperform \textbf{KMeans-Lat}. 
(3) \textbf{NeuroCom} achieves the highest performance in terms of \textbf{sil} and \textbf{nmi} metrics, which shows our model can detect communities with focused topics. This is attributed to the fact that the communities were detected in the same space as the messages were embedded, which preserves semantic similarities between messages. 

\begin{table}[]
\centering
\caption{Community quality evaluation of different models.}
\label{tbl::quantitative-model-comparison}
\vspace{0.2em}
\scriptsize
\begin{tabular}{|c|c|c|c|c|c|c|}
\hline
\multirow{2}{*}{} & \multicolumn{2}{c|}{Support Group} & \multicolumn{2}{|c|}{BGnow} & \multicolumn{2}{|c|}{TuDiabetes} \\ \cline{2-7}
& sil (-ve) & nmi & sil (-ve) & nmi & sil (-ve) & nmi \\ \hline \hline
\textbf{KMeans} & $0.121$ & $0.220$ & $0.140$ & $0.101$ & $0.081$ & $0.352$ \\ \hline
\textbf{KMeans-Lat} & $0.052$ & $0.241$ & $0.073$ & $0.154$ & $0.022$ & $0.411$ \\ \hline
\textbf{Biterm-LDA} & $0.095$ & $0.185$ & $0.082$ & $0.119$ & $0.050$ & $0.322$ \\ \hline
\textbf{RBM} & $0.042$ & $0.287$ & $0.110$ & $0.134$ & $0.031$ & $0.365$ \\ \hline
\textbf{NeuroCom} & $0.039$ & $0.334$ & $0.0166$ & $0.186$ & $0.011$ & $0.570$ \\ \hline
\end{tabular}
\end{table}

\subsection{On Qualitative Comparison of the Model}

It is also intuitive to examine the resulting communities profiles to better understand the output of the \textbf{NeuroCom} model in community profiling and in comparison to existing methods. First, in terms of the number of resulting communities, \textbf{NeuroCom} extracts a higher number of communities, $36$, compared to $22$, $19$, $28$, and $25$ for \textbf{KMeans}, \textbf{KMeans-Lat}, \textbf{Biterm-LDA}, and \textbf{RBM}, respectively. Next, we examined the topics of resulting communities in each method. 
Qualitatively, we find that topics extracted by \textbf{Biterm-LDA} discuss general topics around diabetes, while we can find communities extracted by \textbf{NeuroCom} that are much more focused; each dealing with a specific aspect of the disease. For example \textbf{Biterm-LDA} fails to find small communities such as related to the drug ``Afrezza'' while \textbf{NeuroCom} identifies specific communities such as ``Afrezza'', ``Metformin'' and ``Insulin''.  
These differences are attributed to the fact that traditional approaches work based on word co-occurrences and for such communities they fail to find significant co-occurrence and semantic information. 
We highlight that, as expected, many users are part of multiple communities. From BGnow, 12.6\% of users were in one community, 41.2\% in 2 and 22.7\% in 3 (the rest were in more). For TuDiabetes 11.6\% in 1, 8.4\% in 2 and 26.5\% in 3. Finally in support groups, 28.3\% in 1, 27.2\% in 2 and 21.3\% in 3.

\subsection{Medium Comparisons}

To compare across mediums we can examine inferred demographic profiles of the mediums, as well as the topics of distilled communities. 
Twitter groups were much more skewed, with statistically significant more users inferred to be in the younger age category than in the online forum ($p<0.05$).

We also can compare the topics of communities from TuDiabetes with existing forum categories. All group discussions on TuDiabetes are categorized into the following 14 topics: Community, Type 1 and LADA, TuDiabetes Website, Gestational diabetes, Weight, Type 2, Diabetes Advocacy, Diabetes Complications and other Conditions, Mental and Emotional Wellness, Healthy Living, Diabetes Technology, Food, Treatment, Diabetes and Pregnancy. While interpretation of the topics of communities identified with our method NeuroCom is qualitative, we report that we found several overlaps such as Mental and Emotional Wellness, Food, and Treatment, as shown in Table \ref{tbl::community-profiles}. However given that our method is more organic, and the number of identified communities from NeuroCom, i.e. $36$, is larger, there are some topics that overlap or augment the website categories.

\begin{table}[]
\centering
\footnotesize
\caption{Sample topics and words for each dataset.}
\label{tbl::community-profiles}
\vspace{0.2em}
\scriptsize
\begin{tabular}{|c|c|}
\hline
Community Topic&Top Words \\ \hline
\multicolumn{2}{c}{Support Groups} \\ \hline \hline
medical &disease, stroke, hospitals, patients  \\ \hline
prayer &lord, hope, scripture, God, pray  \\ \hline
exercise & glutes, workout, yoga, getting healthy  \\ \hline
\multicolumn{2}{c}{BGnow} \\ \hline \hline
diet & low carb, lunch, scarifies, recipes, today \\ \hline
Afrezza & Afrezza, inhaler, lung, mankind corp, insulin \\ \hline
running & running, event, run, exercise, daily \\ \hline
\multicolumn{2}{c}{TuDiabetes} \\ \hline \hline
support &empathy, joy, happiness, positivity, feel  \\ \hline
insulin management &insulin, basal, blood, low, dexcom  \\ \hline
diet & carb, sensor, day, weight, meter  \\ \hline
\end{tabular}
\end{table}

\subsection{Community Comparisons Within Mediums}

Assuming community topics are homogeneous across users may be erroneous (e.g. one topic may be more common to, say, women versus men). Further, if community detection is used for recommendation purposes, knowing how various communities are patronized by different types of users can improve  the performance of recommendation. 
Thus here we attempt to profile the resulting communities within each medium.

From a demographic perspective, we found that specific types of communities in TuDiabetes (topics related to insulin management with keywords such as \emph{eat}, \emph{cgm}, \emph{pancreas} and \emph{doctors}) had a statistically significantly proportion of users who were in the older category than the mean proportion ($p<0.05$). As well, of the user profile names that could be linked to gender, more were linked to male (though we caution interpretation of this result due to the large number of user profiles which were not linked to female or male names).
We are also able to describe some qualitative findings from the Twitter datasets. For example, in the support groups, we found that the cluster topic of current events with keywords such as \emph{blackhawks}, \emph{Hillary Clinton}, \emph{primary} also had a statistically significantly proportion of users who were in the older category than the mean proportion ($p<0.05$).

\section{Conclusion}

NeuroCom outperformed existing unsupervised methods based on common cluster evaluation metrics. 
Community detection results showed that topics of distilled communities are interpretable and follow the intuition regarding span of discussion in the Support Group dataset, versus the BGnow and the TuDiabetes forum. Through inferred age categories, we also showed that the online forum had a statistically significantly higher proportion of people in the $<30$ age category. As well, though demographic inference has limitations, there were significantly different proportion of people in the age categories across different communities. While these demographic results are mainly qualitative, we found results that match intuition and can be used in future to improve recommendation approaches or identify concerns of diabetes patients in a more precision and personalized manner. Finally, we compared identified topics to forum categories where available (TuDiabetes), and found that the identified communities in NeuroCom overlap and transcend these existing forum categories.

\section{ Acknowledgments}
This paper was supported in part by grants from the NSF (1643576, 1737987) and NIH (R21AA023901).

\bibliographystyle{aaai}
\footnotesize
\bibliography{www-bibliography} 

\end{document}